# Feature selection and processing of turbulence modeling based on an artificial neural network

Yuhui Yin(尹宇辉),[1] Pu Yang(杨普),[1] Yufei Zhang(张宇飞),[1, a)] Haixin Chen(陈海昕),[1] and Song Fu(符松)[1]

[1] *School of Aerospace Engineering, Tsinghua University, Beijing 100084, China*

## Abstract

Data-driven turbulence modeling has been considered an effective method for improving the prediction accuracy of Reynolds-averaged Navier-Stokes equations. Related studies aimed to solve the discrepancy of traditional turbulence modeling by acquiring specific patterns from high-fidelity data through machine learning methods, such as artificial neural networks. The present study focuses on the unsmoothness and prediction error problems from the aspect of feature selection and processing. The selection criteria for the input features are summarized, and an effective input set is constructed. The effect of the computation grid on the smoothness is studied. A modified feature decomposition method for the spatial orientation feature of the Reynolds stress is proposed. The improved machine learning framework is then applied to the periodic hill database with notably varying geometries. The results of the modified method show significant enhancement in the prediction accuracy and smoothness, including the shape and size of separation areas and the friction and pressure distributions on the wall, which confirms the validity of the approach.

a) Author to whom correspondence should be addressed. Electronic mail: zhangyufei@tsinghua.edu.cn

# I. Introduction

The understanding of the formation and evolution mechanisms of turbulence structures and the accurate prediction of turbulence play important roles in modern engineering design problems. Until now, it has been difficult to solve practical engineering problems by a direct numerical simulation (DNS) or large eddy simulation (LES). The Reynolds-average Navier-Stokes (RANS) equations are still the most commonly used numerical simulation method in aerodynamic design. However, there exist many complex flow phenomena for which the RANS method cannot give satisfactory results, such as flows with swirl, strong pressure gradient, and large mean streamline curvature [1]. Taking the flow separation of aircraft as an example, the stall at large angles of attack, the separation induced by shock wave/boundary layer interaction, and the lift performance deterioration after wing ice accretion all have decisive impacts on flight safety. Unfortunately, researchers have compared several commonly used turbulence models and found that none of them could accurately predict strong adverse pressure gradient separation [2]. There is a consensus that the error in the prediction results is from the discrepancy in Reynolds stress. Therefore, it is necessary to excavate a more accurate model for Reynolds stress, more precisely, a more accurate relationship between Reynolds stress and the mean flow characteristics [3].

The early and commonly used Reynolds stress closure method is a linear eddy-viscosity model (LEVM), which postulates that the Reynolds stress varies linearly with the mean strain rate. The coefficient known as eddy viscosity is then settled through empirical correlations or governing equations. Turbulence models based on LEVM provide advantages, including the robustness of computation, cost savings, and relatively accurate results for attached boundary layer flows. The disadvantages are also obvious. The ignorance of compressibility, the effect of streamline curvature, and the realistic anisotropy characteristic of Reynolds stress cause LEVM to fail to predict complex flows such as the secondary flow [4] in a square pipe and the flow separation of a periodic hill [5]. Strategies aiming to overcome the limitations of LEVM include adding nonlinear terms and directly modeling the Reynolds stress. The former strategy forms nonlinear eddy viscosity models, and the latter forms Reynolds stress transport (RST) models. These improved models can provide a better prediction for certain circumstances. However, the additional empirical terms of the nonlinear eddy-viscosity models can be case-sensitive, and the computation cost and robustness of RST models are poor [6]. In summary, a model that can both give precise results for different types of complex flows and retain robustness and cost savings is still being explored, which leaves a large obstacle for the traditional turbulence modeling framework.

The development of data science and machine learning has brought new perspectives to the fluid mechanics, providing powerful tools for reduced-order modeling, experimental data processing, aerodynamic optimization, turbulence closure, and flow control [7]. Decades of turbulence research have produced numerous high-fidelity turbulence data, including the experimental observations and numerical results of DNS and LES. With the development of computationally efficient statistical inference algorithms and machine learning technologies, large amounts of turbulence data can be utilized more adequately. This framework is referred to as data-driven modeling and is demonstrated in detail in the review [8]. Recent developments in data-driven approaches can be summarized into three categories [8]: (1) Quantifying the uncertainties in RANS models [9]. This approach aims at quantifying the uncertainties in the Reynolds stress tensor [10]-[12] and identifying the regions of high uncertainty [13], which helps bound the uncertainties and clarify the source of the RANS discrepancy. (2) Inferring the discrepancy in the coefficients and the magnitude of terms in the governing equations. This approach selects specific quantities of interest (QoIs) as variables to be inferred in advance. The QoIs can be the model coefficients [14][15] or terms in the governing equations [16]-[18]. These selected variables are then regarded as random fields and inferred to maximize the chosen targets based on the given data, such as the surface pressure distributions [17] or the velocity values at the observation positions [19]. The posterior results of QoIs and corresponding flow field prediction match the targets well if the selection of the QoIs is appropriate enough. (3) Modifying or directly substituting the prediction results of turbulent models using machine learning. This approach aims at constructing a mapping between the QoIs and the mean flow features. Based on massive amounts of training data, machine learning technology can extract multilevel features and give regression or classification results without excessive prior knowledge. The potential of machine learning technology makes it suitable for constructing a mapping between the mean flow and the Reynolds stress of RANS or the subgrid-scale stress of LES [20]. Of the three types of methods demonstrated above, the former two cannot directly provide a predictive model for improving the performance on the "unseen" datasets. However, the machine learning model trained on large datasets can be applied to similar flow cases dominated by the same type of flow phenomenon and flow structure. For example, the train set and test set can be flows over periodic hills with different hill slopes or Reynolds numbers. The generalization capability of the machine learning model makes it more valuable in engineering practice, which can help reduce experimental or DNS simulation costs. Therefore, a RANS model improvement using machine learning is employed in the present research and is discussed in detail.

Turbulence modeling based on machine learning has gained much attention in recent years. Relevant studies can be categorized according to the selection of the training targets. The determination of the training targets reflects the perspectives of the researchers on what the main source of the error is and how to deal with it and are reviewed as follows.

The first view is that the original turbulence model should be kept as a baseline, and the improvement is achieved by modifying some terms in the modeling equations. Tracey et al. [21] used machine learning to rebuild the source terms in the S-A model, which verified the feasibility of the approach. Zhu et al. [25] directly took the eddy viscosity computed by the S-A model as the target to build an artificial neural network. By substituting the original model with the network, the efficiency is improved, and the dependence on grid density decreases. However, if the goal is to improve the original turbulence model, the modified source terms should be first inferred using Bayesian inversion methods and then trained using machine learning methods [22][23]. More recent research combined the inference and machine learning into a single process, which directly inferred the weights and biases [24]. In these studies, the modification is realized by multiplying a factor by the production or dissipation term, which varies in the flow field. Because the effect on the Reynolds stress is propagated through the transport equations, it is easier to achieve the smoothness requirement of the modified Reynolds stress field. However, the eddy-viscosity hypothesis bounds the anisotropy of the Reynolds stress. There still exists a difference between the theoretical optimum of the current view and the true Reynolds stress.

The second view is that the whole Reynolds stress should be selected as the training target. In Ling's work [26], the Reynolds stress is represented as the linear combination of ten integrity bases formed by the mean strain rate tensor and mean rotation rate tensor. The output variables of the neural network are the coefficients that are then combined with corresponding tensor bases to form the predicted Reynolds stress anisotropic tensor. The idea of predicting the coefficients of nonlinear turbulent viscosity models inspired some later studies [27]-[29]. In addition to the linear combination, eigendecomposition is also employed. As the Reynolds stress is a symmetric second-order tensor, the eigendecomposition gives out real-orthogonal eigenvalues and eigenvectors. The eigenvectors can be described by Euler's angles [30][31] or quaternions [32]. Applying the decomposition to both the true Reynolds stress and the RANS predicted Reynolds stress, the discrepancy field for each feature can be obtained and taken as the targets of machine learning. Regardless of which method is employed, the current view overcomes the deficiency in the former view, which neglects the anisotropy modification. Specifically, the eigendecomposition method gives rational feature fields with clear physical meaning, which gives a better generalization between the different flow cases. Therefore, the eigendecomposition method is employed in the following study.

The construction of the prediction framework in this paper mainly follows the work of Wu et al. [32]; therefore, it is not the focus of this paper. The main problem that we are concerned with is the unsmoothness and prediction error of the output results, focusing on feature selection and processing for both the input and output. From the aspect of input features, the selection criteria based on tensor analysis and flow structure identification are summarized. The input feature set is constructed accordingly. In addition, the effect of the computational grid on the input features is also investigated. Then, from the aspect of output features, a modified representation method of the eigenvectors or spatial direction based on Euler's angles is proposed. The method is shown to behave better in decreasing the unsmoothness of the true values, which is beneficial for the training and predicting processes. A set of flows over periodic hills with notably different slopes is selected as the test case to further verify the corrections. The final results prove the performance improvement.

## II. Methodology

### A. The framework of turbulence modeling based on machine learning

Essential statements of the turbulence modeling framework based on machine learning are needed to illustrate the following feature processing methods. The three-dimensional compressible RANS equations for a Newtonian fluid without heat transfer and heat generation are

$$
\begin{aligned}
&\frac{\partial \rho}{\partial t} + \nabla \cdot (\rho \mathbf{u}) = 0 \\
&\frac{\partial}{\partial t}(\rho \mathbf{u}) + \nabla \cdot (\rho \mathbf{u}\mathbf{u}) = \rho \mathbf{f} + \nabla \cdot \mathbf{T} \\
&\frac{\partial}{\partial t}(\rho e) + \nabla \cdot (\rho e \mathbf{u}) = \nabla \cdot (\mathbf{T} \cdot \mathbf{u}) + \rho \mathbf{f} \cdot \mathbf{u} \\
&\mathbf{T} = -p\mathbf{I} + \left[ \lambda (\nabla \cdot \mathbf{u})\mathbf{I} + \mu (\nabla \mathbf{u} + \mathbf{u}\nabla) \right] + \boldsymbol{\tau}
\end{aligned}
\tag{1}
$$

where $\rho$, $\mathbf{u}$, $\mathbf{f}$, $e$, $p$, $\lambda$, and $\mu$ are the mean density, mean velocity, body force, mean total energy, mean pressure, bulk viscosity, and molecular viscosity, respectively. $\mathbf{T}$ represents the total stress tensor, including the pressure, the molecular viscous stress, and Reynolds stress $\boldsymbol{\tau}$. In traditional turbulence modeling, the Reynolds stress is determined by turbulence model equations. The solving process reaches the end when both the RANS equations and turbulence model equations converge.

The role of machine learning in turbulence modeling is to provide a predictive model that provides the Reynolds stress with certain input features, more precisely, a regression function. The function is first built based on the existing DNS database, which is referred to as training in machine learning techniques. Then, the function can be applied to new unseen flows (referred to as the test case). The selection of the prediction targets, input and output features, and machine learning algorithm all have decisive effects on the performance.

The aim of the machine learning framework in the present study is to predict the Reynolds stress difference between the true values given by the DNS results and baseline values given by the linear eddy viscosity model; that is to say, the prediction target is $\Delta\boldsymbol{\tau} = \boldsymbol{\tau}^{\mathrm{DNS}} - \boldsymbol{\tau}^{\mathrm{RANS}}$. Using the difference value rather than the full value, the model can better utilize the RANS result when the difference is not significant. The regression only needs to focus on the large discrepancy, which decreases the training difficulty. The other benefit is that the coordinate invariance requirement of the output features can be automatically fulfilled[30]. Therefore, the discrepancy field of Reynolds stress is employed in several works as the prediction target [30][32]. In addition, the type of prediction target also determines the source of the input information and the coupling method between machine learning and the mean flow solution. In the present study, when applying the trained model to the test case, the $\Delta\boldsymbol{\tau}$ predicted by machine learning combined with the $\boldsymbol{\tau}^{\mathrm{RANS}}$ directly provides the predicted $\boldsymbol{\tau}^{\mathrm{DNS}}$ of the test case. To further obtain the modified mean flow field, the predicted Reynolds stress has to be substituted into the original value and kept constant during the iteration of the RANS equations until the residual reconverges, which means that there is a loose coupling in which the machine learning model is only called once before the second round of iteration starts. This also leads to the input information for the prediction model only coming from the RANS result. As a comparison, if the prediction target is the direct true value of the Reynolds stress and the baseline turbulence model equation is abandoned [25], the input information will come from the true mean flow field and the machine learning model has to be called in every iteration during the solving process to guarantee the correct convergence, which is tight coupling.

The entire procedure of the turbulence modeling framework based on machine learning is summarized as follows and is shown in Fig. 1.

(1) Generate computational grids for both the training set and test set and run the baseline RANS simulation.

(2) Obtain the mean flow features $\mathbf{q}$ from the mean flow quantities $\mathbf{x}$: $\mathbf{q}(\mathbf{x})\big|_{\mathrm{train}}$ and the Reynolds stress of RANS simulation on the train set: $\boldsymbol{\tau}^{\mathrm{RANS}}\big|_{\mathrm{train}}$.

(3) Interpolate the DNS results to the RANS computational grids to obtain $\boldsymbol{\tau}^{\mathrm{DNS}}\big|_{\mathrm{train}}$ and the discrepancy field $\Delta\boldsymbol{\tau}\big|_{\mathrm{train}} = \boldsymbol{\tau}^{\mathrm{DNS}}\big|_{\mathrm{train}} - \boldsymbol{\tau}^{\mathrm{RANS}}\big|_{\mathrm{train}}$.

(4) Construct the mapping function $f : \mathbf{q} \mapsto \Delta\boldsymbol{\tau}$ based on the training set. First, parameter tuning is performed using cross-validation, and then the neural network is trained using optimized hyperparameters.

(5) Apply the neural network to predict the discrepancy field $\Delta\boldsymbol{\tau}\big|_{\mathrm{test}}$ using mean flow features $\mathbf{q}\big|_{\mathrm{test}}$ on the test set and combine the discrepancy field with the Reynolds stress of RANS simulation $\boldsymbol{\tau}^{\mathrm{RANS}}\big|_{\mathrm{test}}$ to obtain predicted results $\boldsymbol{\tau}^{\mathrm{pre}}\big|_{\mathrm{test}}$.

(6) Substitute $\boldsymbol{\tau}^{\mathrm{pre}}\big|_{\mathrm{test}}$ into the CFD solver and freeze, solve the RANS equation until reconvergence to obtain the final corrected mean flow results $\mathbf{x}'\big|_{\mathrm{test}}$.

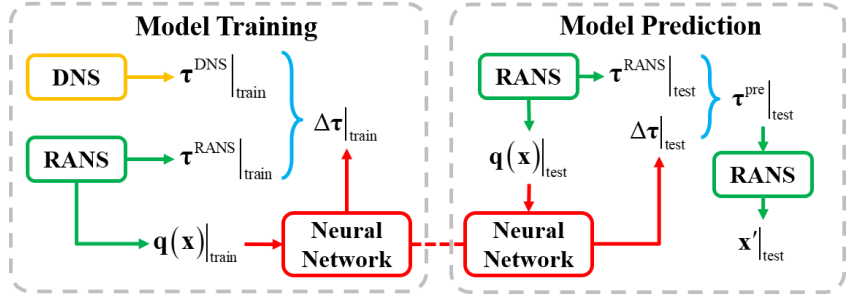

Fig. 1 Schematic of the machine learning framework

## B. Input feature selection based on tensor and physical analysis

As illustrated in Sec. II A, the input features in this work should come from the RANS predicted mean flow quantities. There are three basic principles during construction:

(1) The completeness. The input feature set should include all the possible information that is relevant to the Reynolds stress distribution.

(2) The compactness. Invalid or redundant information should be identified and removed.

(3) The realizability. Selected features should be consistently effective under various circumstances, such as different reference frames or flow directions, as long as the flow structure remains unchanged. The realizability can be restated from a tensor perspective: the Reynolds stress should be an isotropic tensor function of its relied scalars, vectors, and tensors.

The first two properties help improve the performance of the trained neural network, making it more extrapolatable and accurate. The last property ensures that the data-drive model is reasonable, as in traditional turbulence modeling. Under these constraints, researchers have made considerable efforts to choose the proper mean flow features [33]-[35]. By summarizing the previous works, two selection criteria based separately on tensor analysis and characteristic identification are proposed in this paper. The features appearing in previous studies can be categorized into these two considerations, and some new effective features are constructed and added to the input set under the guidance of the criteria.

## 1. Reynolds stress representation: a tensor analysis perspective

The first consideration is to search for the tensors or vectors that the Reynolds stress depends on and to construct a complete tensor basis for representing the Reynolds stress. The nonlinear eddy-viscosity hypothesis, as an example, is a typical representation method of turbulence modeling, which assumes the Reynolds stress as a combination of polynomials formed by the mean flow strain rate $\mathbf{S}$ and the rotation rate $\mathbf{\Omega}$. Pope [36] used the tensor function representation theory to derive an integrity basis with 10 components to obtain the compact form of the polynomials. The integrity basis guarantees that every symmetric deviatoric second-order tensor formed by $\mathbf{S}$ and $\mathbf{\Omega}$ can be expressed as a linear combination of these 10 tensors. Inspired by this thought, Wu et al. [32] expanded the dependence of the Reynolds stress by adding pressure gradient $\nabla p$ and turbulent kinetic energy (TKE) gradient $\nabla k$ to supplement the effect of strong pressure changes and the turbulent nonequilibrium effect. The two gradients are transformed into corresponding antisymmetric tensors $\mathbf{A}_p$ and $\mathbf{A}_k$ for convenience. The integrity basis with 47 components formed by these four tensors or vectors is similarly derived. Because the integrity basis components are tensors and cannot be directly used as input features, Wu et al. [32] chose the first invariant, i.e., the trace of each tensor as the input features.

The 47 features above have been proven to work well in machine learning and therefore are employed as the input features. In addition, the same normalization method is also used to process the four raw tensors [32]. The number of the feature set can be reduced to 17 in the two-dimensional circumstance after removing the zero values and repetitive characters with the same absolute values, as shown below. The 17 features are numbered from $q_1$ to $q_{17}$. (The $|\cdot|$ denotes the vector norm, and the $\|\cdot\|$ denotes the Frobenius matrix norm. The variable $\varepsilon$ is the turbulent dissipation rate.)

$$
\begin{array}{ccc}
(n_{\mathrm{sym}}, n_{\mathrm{anti}}) & \text{Feature No.} & \text{Invariants} \\
(1,0) & q_1 & \hat{\mathbf{S}}^2 \\
(0,1) & q_2 - q_3 & \hat{\mathbf{A}}_p^2, \hat{\mathbf{A}}_k^2 \\
(1,1) & q_4 - q_8 & \hat{\mathbf{\Omega}}^2\hat{\mathbf{S}}^2, \hat{\mathbf{A}}_p^2\hat{\mathbf{S}}, \hat{\mathbf{A}}_p^2\hat{\mathbf{S}}^2, \hat{\mathbf{A}}_k^2\hat{\mathbf{S}}, \hat{\mathbf{A}}_k^2\hat{\mathbf{S}}^2 \\
(0,2) & q_9 & \hat{\mathbf{A}}_p\hat{\mathbf{A}}_k \\
(1,2) & q_{10} - q_{13} & \hat{\mathbf{A}}_p^2\hat{\mathbf{\Omega}}\hat{\mathbf{S}}, \hat{\mathbf{A}}_k^2\hat{\mathbf{\Omega}}\hat{\mathbf{S}}, \hat{\mathbf{A}}_p\hat{\mathbf{A}}_k\hat{\mathbf{S}}, \hat{\mathbf{A}}_p\hat{\mathbf{A}}_k\hat{\mathbf{S}}^2 \\
(0,3) & q_{14} & \hat{\mathbf{\Omega}}\hat{\mathbf{A}}_p\hat{\mathbf{A}}_k \\
(1,3) & q_{15} - q_{17} & \hat{\mathbf{\Omega}}\hat{\mathbf{A}}_p\hat{\mathbf{A}}_k\hat{\mathbf{S}}, \hat{\mathbf{\Omega}}\hat{\mathbf{A}}_p\hat{\mathbf{A}}_k\hat{\mathbf{S}}^2, \hat{\mathbf{\Omega}}\hat{\mathbf{A}}_k\hat{\mathbf{A}}_p\hat{\mathbf{S}}^2
\end{array}
\tag{2}
$$

where $\hat{\mathbf{S}} = \dfrac{\mathbf{S}}{\|\mathbf{S}\| + \varepsilon/k}, \hat{\mathbf{\Omega}} = \dfrac{\mathbf{\Omega}}{2\|\mathbf{\Omega}\|}, \hat{\mathbf{A}}_p = \dfrac{\mathbf{A}_p}{\|\mathbf{A}_p\| + \rho|(\mathbf{u}\cdot\nabla)\mathbf{u}|}, \hat{\mathbf{A}}_k = \dfrac{\mathbf{A}_k}{\|\mathbf{A}_k\| + \varepsilon/\sqrt{k}}$. $n_{\mathrm{sym}}$ and $n_{\mathrm{anti}}$ refer to the numbers of symmetric/antisymmetric tensors employed in the invariants.

However, there exists one problem: whether all the characteristics of the Reynolds stress can be expressed only by the combination of invariances. As stated above, the eigendecomposition of the Reynolds stress provides three eigenvalues (TKE and two anisotropy features) and three eigenvectors. The TKE is theoretically independent of the 47 features because it represents the spherical part of the Reynolds stress and the integrity basis represents the deviatoric part. The three eigenvectors are not invariant and therefore cannot be fully represented by the invariance features. By taking the Reynolds stress discrepancy as the target, the magnitude problem of TKE and the invariance problem of the eigenvectors can be partially solved, which is illustrated in Sec. II C. However, the dependence problems are not taken into account. Therefore, it is necessary to investigate other features as supplements.

## 2. Discrepancy detection and measurement: a characteristic identification perspective

The second consideration is to search for the features by detecting and measuring the discrepancy between the RANS predicted flow field and the true DNS flow field. Compared with the representation method above, finding the markers to identify the difference and regarding them as inputs are more straight forward and relevant to the target of machine learning. In addition to the marking function, the selected features should have clear physical meanings to guarantee a similar performance in the flow with similar flow structures. There are two methods used in constructing the markers: (1) criteria identifying the key flow structures; (2) ratios reflecting the relative magnitude, which are illustrated as follows.

**(1) Criteria identifying the key flow structures.** A discrepancy generally occurs with the complex flow structures. Taking the flow over a periodic hill as an example, the error of the RANS prediction mainly comes from the prediction of the shear layer, strong adverse pressure gradient, and swirl flow in the separation bubble. The insufficient modeling of the mixing effect is a problem in RANS simulation [6]. In the traditional turbulence modeling framework, the baseline RANS model is modified by varying the magnitude of the turbulence production or dissipation in the concerned areas [37][38]. The functions used in these studies to locate the concerned areas can also be used in a machine learning framework as input features, which have clear physical meanings and desired invariance properties. Three markers are selected in the present research and are listed as follows.

- The marker of the shear layer and swirl flow [16]: $f_1 = d^2 \Omega / (\nu_t + \nu)$, where $d$ is the distance to the wall, $\nu_t$ is the RANS predicted eddy viscosity, and $\Omega = \sqrt{2\Omega_{ij}\Omega_{ij}}$.

- The marker of the adverse pressure gradient [13]:

$$f_2 = \frac{\hat{\mathbf{u}} \cdot (\nabla p)}{1 + \left| \hat{\mathbf{u}} \cdot (\nabla p) \right|}, \text{ where } \hat{\mathbf{u}} = \frac{\mathbf{u}}{\sqrt{\mathbf{u} \cdot \mathbf{u}}}, (\nabla p) = \frac{\nabla p}{\sqrt{(\nabla p) \cdot (\nabla p)}}.$$

- The marker of the boundary layer [39]: $f_3 = 1 - \tanh\left(\left[8 r_d\right]^3\right)$, where $r_d = \dfrac{\nu_t + \nu}{\kappa^2 d^2 \sqrt{u_{i,j} u_{i,j}}}$.

**(2) Ratios reflecting the relative magnitude.** The Reynolds stress discrepancy of the RANS prediction and DNS results can also be investigated from the relative magnitude of the turbulence quantities. For example, the nonequilibrium effect can be measured by the ratio of the production term and dissipation term; the relative importance of turbulent inertia to the viscosity can be measured by the turbulent Reynolds number. The ratio features are naturally normalized because of the definition and therefore beneficial for the generalization capacity. Five ratios are selected in this work and are listed as follows.

- The ratio of the turbulent kinetic energy to the mean kinetic energy: $f_4 = 2k / \left(u^2 + v^2 + w^2\right)$

- The ratio of the production term to the dissipation term in the TKE equation: $f_5 = P_k / \varepsilon$.

- The wall-distance-based turbulent Reynolds number [40]: $f_6 = \min\left(\dfrac{\sqrt{k}d}{50\nu}, 2\right)$.

- The ratio of the turbulent time scale to the mean flow time scale: $f_7 = Sk / \varepsilon$.

- The ratio of the turbulent viscosity to the mean flow viscosity: $f_8 = \nu_t / \nu$.

In summary, there are 25 normalized-invariant features constructed from the mean flow variables and selected as the input features for the machine learning framework. The feature importance of the selected features is verified using the random forest technique in Sec. III B, especially the newly added features based on characteristic identification.

## C. Improvement of the Reynolds stress discrepancy representation

Reynolds stress discrepancy $\Delta\boldsymbol{\tau} = \boldsymbol{\tau}^{\mathrm{DNS}} - \boldsymbol{\tau}^{\mathrm{RANS}}$ is the target of machine learning. However, the components of the tensor cannot be directly used as the output features and must be converted to features that satisfy the following principles:

(1) The degree of freedom should remain the same as the number of independent components, which is six for the Reynolds stress of three-dimensional flow and four for two-dimensional flow;

(2) The features should be invariant to rotations in the reference frame to remain consistent

with the input features.

The eigendecomposition method has been commonly used for the Reynolds stress representation in machine learning [10][30][32], as shown in Eq. (3).

$$\boldsymbol{\tau} = 2k\left(\frac{1}{3}\mathbf{I} + \mathbf{b}\right) = 2k\left(\frac{1}{3}\mathbf{I} + \mathbf{V}\boldsymbol{\Lambda}\mathbf{V}^{\mathrm{T}}\right) \tag{3}$$

where $k$ is the TKE; $\mathbf{I}$ is the second-order identity tensor; $\mathbf{b}$ is the normalized Reynolds stress deviatoric tensor; $\mathbf{V} = [\mathbf{v}_1, \mathbf{v}_2, \mathbf{v}_3]$ is the orthonormal matrix constructed by three eigenvectors; $\boldsymbol{\Lambda} = \mathrm{diag}[\lambda_1, \lambda_2, \lambda_3]$ is the diagonal matrix constructed by three eigenvalues that are ranked from large to small and satisfy $\lambda_1 + \lambda_2 + \lambda_3 = 0$. Because of the realizability of the Reynolds stress (normal stress being nonnegative), the value range of $\lambda_i$ is constrained in a feasible region. The barycentric coordinate [41] provides a linear projection from the feasible region of the eigenvalues to the physically realizable turbulence states, which is similar to the Lumley triangle. The eigenvalues are first converted to triangular coordinates $[C_1, C_2, C_3]$ and then projected to Cartesian coordinates $[\xi, \eta]$, as shown in Eq. (4).

$$\begin{cases} C_1 = \lambda_1 - \lambda_2 \\ C_2 = 2(\lambda_2 - \lambda_3) \\ C_3 = 3\lambda_3 + 1 \end{cases} \Rightarrow \begin{cases} \xi = X_{1C}C_1 + X_{2C}C_2 + X_{3C}C_3 \\ \eta = Y_{1C}C_1 + Y_{2C}C_2 + Y_{3C}C_3 \end{cases} \tag{4}$$

In summary, the magnitude of TKE $k$ and the shape features $(\xi, \eta)$ are acquired by eigendecomposition. Because they are naturally invariant, the difference values are still invariant and can be used as output features. Specifically, the TKE discrepancy is characterized by the logarithm of the ratio of $k^{\mathrm{DNS}}$ to $k^{\mathrm{RANS}}$ to guarantee that the predicted TKE is nonnegative, as shown in Eq. (5).

$$\Delta\log k \equiv \log\frac{k^{\mathrm{DNS}}}{k^{\mathrm{RANS}}}, \quad \Delta\xi \equiv \xi^{\mathrm{DNS}} - \xi^{\mathrm{RANS}}, \quad \Delta\eta \equiv \eta^{\mathrm{DNS}} - \eta^{\mathrm{RANS}} \tag{5}$$

Different from the features above, the spatial orientations themselves are not invariant. However, the orientation discrepancy of $\boldsymbol{\tau}^{\mathrm{DNS}}$ and $\boldsymbol{\tau}^{\mathrm{RANS}}$ can be processed to invariants by directly rotating the principal coordinate of $\tau^{\text{RANS}}$ to the principal coordinate of $\tau^{\text{DNS}}$. By switching the reference coordinate to the RANS predicted Reynolds stress principal coordinate, the orientation difference is independent of the global coordinate selection. In addition to supplying formal invariance, the treatment also has a clear physical meaning, indicating a rational and extrapolatable distribution rule of the discrepancy. There are two methods used to measure the rotation: the Euler angle and the unit quaternion [42]. Both can realize an arbitrary rotation in a three-dimensional space. In the two-dimensional flow case, the rotation axis is always the z-axis normal to the flow plane, and the two methods are equivalent, which also means that they face the same problem of the spatial step distribution.

The step distribution problem results from the conflict between the periodicity of the rotation angle and the single-valued requirement of the regression function. The rotation between two Reynolds stresses can be analogized with the rotation between two ellipsoids because the orientation is bidirectional. Therefore, the period of the rotation angle is reduced to $\pi$. The starting point of the semicircle is arbitrary. In this work, it is set as $[-\pi/2, \pi/2)$. Once the range is set, a step value appears, as shown in Fig. 2. The two rotations are separately marked by a red line and a blue line corresponding to $\theta_1$ and $\theta_2$. The absolute values of both angles are close to $\pi/2$, but one is positive and the other is negative. It can be observed that the two rotations are quite similar and probably close to each other in the flow field. However, because of the separation of the value range, $\theta_1$ is close to $\pi/2$ and $\theta_2$ is close to $-\pi/2$, showing a great numerical difference.

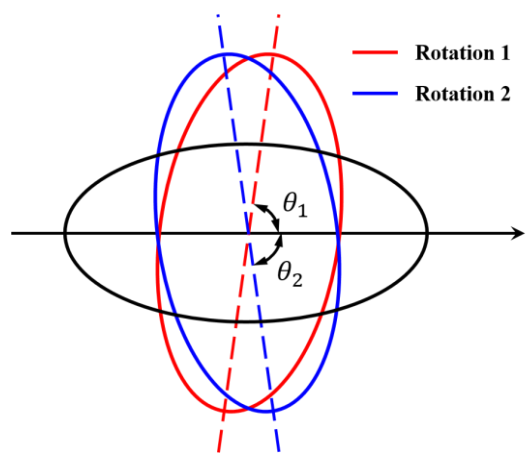

Fig. 2 Illustration of the step distribution problem.

Several issues need to be discussed regarding the step distribution. First, it is an inherent problem regardless of which starting point is selected. For example, if the semicircle is set as $[0, \pi)$, rotation angles near 0 or $\pi$ (spatially equivalent) will face the same phenomenon. Second, the unit quaternion method cannot solve the problem either; it only changes the step value from $\pm\pi/2$ to $\pm\pi/4$ because the unit quaternion uses $\theta/2$ in its expression. Third, although the problem does not directly affect the computation of the rotation and reconstruction of the Reynolds stress, it brings a large error to the training of the regression function, as shown in Fig. 3 (a). The abscissa represents a spatially continuous distribution, such as a cross-section in the flow field. The black solid line is an original distribution of rotation angle $\theta$. There exists a step value from $\pi/2$ to $-\pi/2$. As stated above, this is a numerical phenomenon. However, the regression function constructed by the neural network is a smooth distribution function. A large error appears near the step value position, which is indicated by the red dashed line.

To overcome the numerical discontinuity of the original distribution, an additional feature decomposition is proposed in this work. Rotation angle $\theta$ is decomposed to a cosine value, $\cos(\theta)$ and a sign value, $\text{sgn}(\theta)$, respectively shown in Fig. 3 (b) and (c). It can be observed that the discontinuity no longer exists in the $\cos(\theta)$ distribution because of the property of the cosine function. Such a smooth distribution is quite suitable for neural network regression. In addition to numerical convenience, the cosine value directly appears in the rotation matrix component and therefore is physically interpretable. After removing the magnitude information, $\text{sgn}(\theta)$ can be well predicted using a classification network, which is also a typical application scenario of the artificial neural network. By separating the magnitude and sign, the two features can both achieve better prediction performance compared with the combined feature.

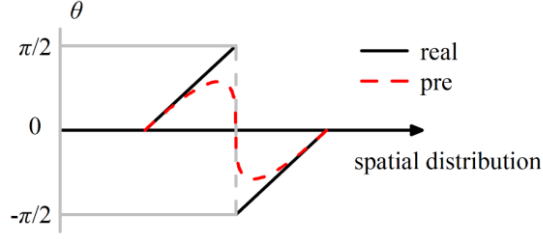

(a) The original rotation angle, $\theta$

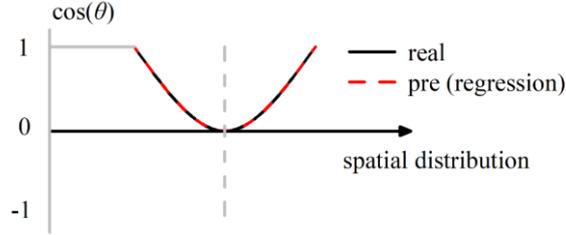

(b) The first decomposed feature: $\cos(\theta)$

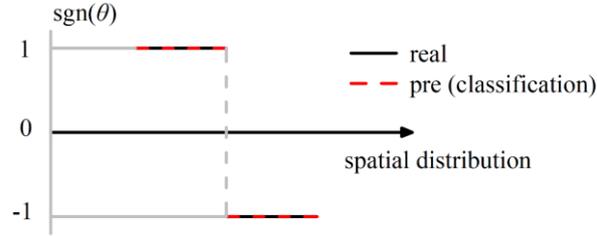

(c) The second decomposed feature: $\mathrm{sgn}(\theta)$

Fig. 3 Comparison of the real and predicted distributions of spatial orientation.

In summary, the Euler angle method of the orders $\Delta\theta_z^1$, $\Delta\theta_x^2$, and $\Delta\theta_z^3$ is employed to represent the orientation difference between the baseline and the true Reynolds stress. The additional decomposition above increases the total number of output features to nine for the three-dimensional flow. For the two-dimensional flow, the last orientation information ($\Delta\theta_x^2$, $\Delta\theta_z^3$) is omitted. Therefore, five features are finally settled, as shown below. In the machine learning process, one network is trained for each feature. The sign feature is trained as a classifier, and the others are trained as regressors.

$$\left\{ \Delta\log k, \Delta\xi, \Delta\eta, \cos\left(\Delta\theta_z^1\right), \mathrm{sgn}\left(\Delta\theta_z^1\right) \right\} \tag{6}$$

## D. Machine learning algorithm and parameter optimization

The machine learning algorithm used in the present work is the artificial neural network (ANN) implemented in Python based on the Keras deep learning package [43]. Inspired by the biological neural network of the neurons in a brain, ANN is based on a collection of connected units or nodes called artificial neurons. The input data are transferred into those neurons, summed up, computed by a nonlinear activation function, and then sent to other neurons. ANNs are famous for their powerful nonlinear fitting capacity and are commonly used in many research fields. Therefore, ANNs are suitable for constructing a mapping of the complex relationship between the mean flow quantities and the turbulence quantities. However, one notable disadvantage of ANNs is a large set of hyperparameters that require tuning. The selection of the number of hidden layers, neuron number of each layer, and training epochs all affect the results, especially in turbulence modeling, which does not have enough empirical evidence. Therefore, the optimization of hyperparameters based on cross-validation [44] is employed.

Cross-validation has been commonly used for assessing how the prediction results are generalized to an unknown dataset. Parameter tuning based on cross-validation can effectively avoid the overfitting problem. The entire dataset is first separated into the training set, validation set, and test set. A set of predictive models with different hyperparameters is trained on the training set and applied to the validation set. The performance of the validation set shows the generalization capacity and can be used to choose the most proper model. Then, the best model is applied to the test set to acquire the final prediction results. In this work, k-fold cross-validation is employed. The datasets, except for the test set, are separated into $K$ groups. For one hyperparameter set, each group is selected as the validation set once, and the other ($K$-1) groups are training sets, generating $K$ predictive models and corresponding error on the validation set. The final cross-validation error is the mean squared error (MSE) of the $K$ models. This is realized using the scikit-learn package in Python [45].

# III. Results

The flow over periodic hills is investigated to verify the method proposed in this paper. As a representative of a flow with massive separation, it is difficult for the RANS method to predict the flow accurately. In addition, there have been many DNS and LES results presented, providing enough datasets for training and testing. The computation program with the Reynolds stress substitution function is developed. The true Reynolds stress field is substituted into the program to acquire a reconverged mean flow field. The mean flow field is compared with the true mean flow to verify the correctness and to analyze the sensitivity of such an explicit treatment. Then, improvements in the feature relevance and the smoothness of the input features and output features are presented, including the effect of computation grid smoothing, the effect of the newly added features, and the effect of the improved representation of spatial orientation. Finally, the ANN based on the proposed method is trained. The generalization performance on massively varying geometries is presented.

## A. Problem setup and theoretical optimum result

A schematic of the flow over periodic hills is shown in Fig. 4 (a). The geometrical parameters of the periodic hill include the distance between two hill crests $L_x$, the distance between the valley and the top boundary $L_y$, crest height $H$, and hill width $W$. The baseline parameters satisfy $L_x/H =$ 9, $L_y/H = 3.036$, and $W/H = 1.929$. The flow enters the domain from the left side, and the right side is connected with the left side to achieve periodicity. The bottom and top are all set as no-slip solid walls. The flow is driven by a uniform body force. In order to match the flow condition with the DNS result, the body force is adjusted to acquire the settled mass flow target. The detailed steps are listed as follow:

(1) Determine the mass flow target that can fulfill the mean flow Reynolds number based on the bulk velocity at the crest.

(2) Because the left and right sides of the periodic hill are set as periodic conditions and connected, there is no pressure gradient. Therefore, a uniform body force is added to each grid cell in solving to drive the flow.

(3) During the iteration of the CFD solving process, the body force is firstly kept unchanged for several steps to build the initial field and avoid the divergence. After the initial field is built, the mass flow at the crest is computed after each iteration. Then the body force is adjusted according to that the current mass flow is smaller or larger than the target mass flow. A proportional & differential (PD) control method is adopted in the program to increase the convergence speed and avoid the overshoot during the adjustment.

(4) When the solving process converges, the body force converges to a certain value and the mass flow converges to the target settled above.

Parameterized periodic hill geometries with different steepness are investigated in this work to verify the generalization capacity on different geometries, as shown in Fig. 4 (b). The geometries and DNS results are provided in the work of Xiao et al. [46]. Factor $\alpha$ is introduced to scale the hill width by multiplying the width by $\alpha$. Therefore, the distance in the flow direction is $L_x/H = 3.858\alpha + 5.142$. The original geometry corresponds to $\alpha = 1.0$. Five geometries with $\alpha$ varying from 0.5 to 1.5 are selected. In the present study, $\alpha = 0.8$ and 1.2 cases construct the training set, and $\alpha = 0.5$, 1.0, and 1.5 cases construct the test set. Therefore, the prediction of the test set includes interpolation and extrapolation on the geometrical level. The Reynolds number based on $H$ and the bulk velocity at crest $U_b$ is $\text{Re}_b = 5600$. The DNS simulations are under the incompressible condition [46].

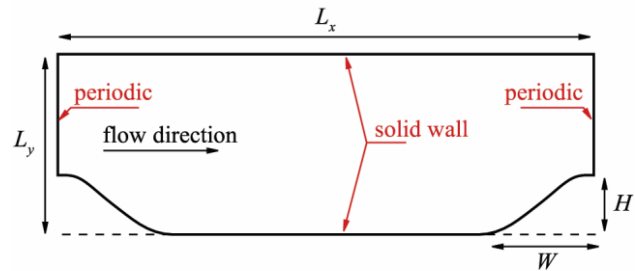

(a) The geometrical parameters and boundary conditions

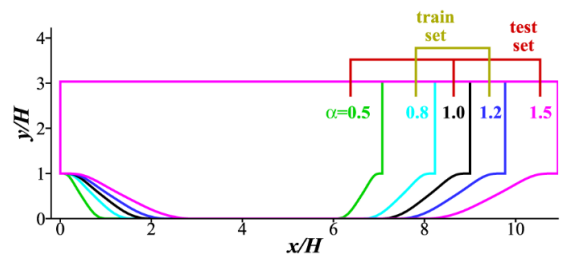

(b) Parameterized periodic hill geometries with different steepness

Fig. 4 Schematic of the flow over periodic hills

The compressible code CFL3D version 6.7 is used as the CFD solver for the following computations [47]. The solver is needed not only in performing baseline RANS simulation but also in acquiring the mean flow field with a given Reynolds stress. The Mach number based on $U_b$ is set to 0.2 to minimize the compressibility. First, the baseline RANS simulations are performed for the five cases above. The turbulence model used is Menter's $k$-$\omega$ SST model. The flow field results all have good convergence, but the separation area and the reattachment location clearly differ from those in the DNS results.

In addition to acquiring the baseline RANS results, the mean flow fields based on the Reynolds stress from the DNS results are also computed. The detailed steps are listed as follows:

(1) The DNS database provides the averaged second-order correlation fields of the fluctuating velocity, which is the Reynolds stress field on the DNS grids according to the definition.

(2) We interpolate the Reynolds stress fields on fine grids of DNS onto the RANS coarse grids, acquiring the fields that have one-to-one correspondences with the RANS predicted mean flow features in the spatial distribution.

(3) The Reynolds stress fields on RANS coarse grids are then substituted into the CFD solver and keep frozen during the iteration. The original SST predicted Reynolds stress fields are replaced. When the solving process converges, the mean flow results correspond to the "theoretical optimum" results.

It can be found that the results are acquired using the true value of the averaged Reynolds stress field from DNS results. If a machine learning model can completely understand the DNS behavior which means that the model has reached its upper bound or theoretical optimum, the output will be the same with the Reynolds stress we used above. Therefore, the results are referred to as "theoretical optimum" in this work.

The comparison of the Reynolds stresses and the mean velocities between the RANS result, the DNS result, and the theoretical optimum result is shown in Fig. 5, taking the $\alpha = 1.0$ case for illustration. The Reynolds stress field of the theoretical optimum is the same as the DNS result and therefore omitted. The TKE and the Reynolds shear stress discrepancy between the RANS and DNS are mainly located near the shear layer and the separation area, as shown in Fig. 5 (a) and (b). The RANS result underestimates the turbulence production, leading to a larger separation area and delayed reattachment, as shown in Fig. 5 (c).

The comparison of the velocity profiles between the DNS result (black solid lines) and the theoretical optimum result (red dashed lines) is also shown in Fig. 5 (c). Although the Reynolds stress field is kept the same, the interpolation from the DNS grid to the RANS grid may induce a numerical error. In addition, the number of grid cells, the spatial discretization accuracy, and the time advancement method are also different. These differences finally lead to the discrepancy of the mean flow result, which is referred to as the sensitivity of the solved quantities to the modeled terms [48]. In the present situation, the sensitivity problem is not notable, as shown in Fig. 5 (c). The separation area, recirculation velocity profile, and reattachment location are almost the same. Some relatively obvious discrepancy is located near the top wall. The result gives an upper limit of the following work. This indicates that consistency in the mean velocity results will be achieved if the predicted Reynolds stress field is sufficiently accurate.

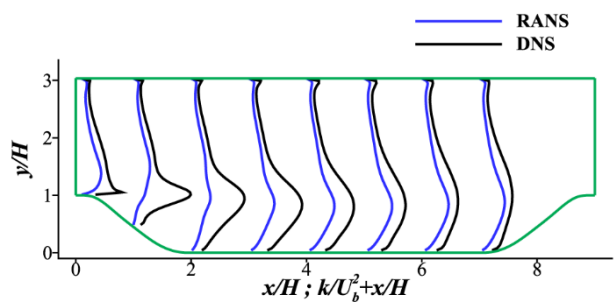

(a) TKE profiles

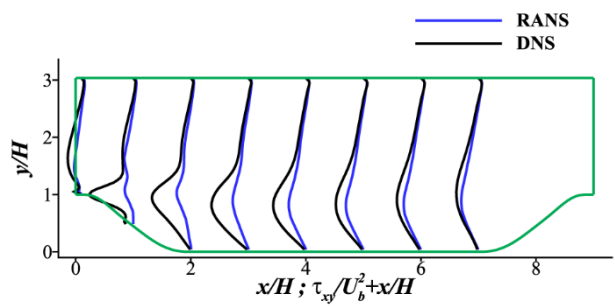

(b) Reynolds shear stress profiles

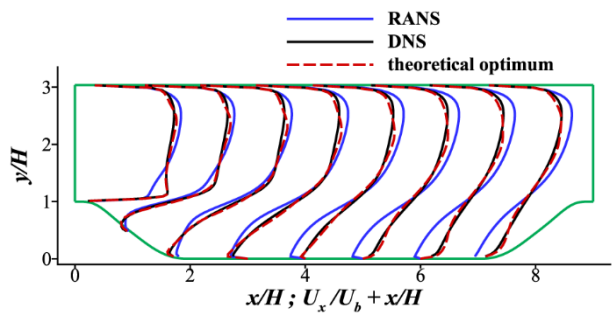

(c) Mean velocity profiles

Fig. 5 Comparison of RANS, DNS, and theoretical optimum results for $\alpha = 1.0$

## B. Improvement in the feature relevance and smoothness

The feature relevance between the input and output directly affects the generalization capacity of the ANN. If the input features are poorly relevant to the targets, the strong fitting capacity will still decrease the error for the training set. However, the prediction error for the test set will be large because no physical and rational relationship is learned. Meanwhile, the spatial smoothness of the input and output features also has significant impacts on the performance of the neural network. The flow field in the case without shock-wave or multiphase flow is continuous. However, the numerical processes may lead to an unexpected discontinuity. Because these phenomena are unphysical, they will surely harm the extraction of a realistic physical correlation and therefore need to be avoided. To improve the two aspects above, three efforts are taken in this work. The first is to test the effect of the computational grid on the input feature smoothness. The second is to verify the effect of the newly added features based on characteristic identification. The third is to investigate the effect of the new process of spatial orientation on the output feature smoothness.

### 1. The effect of the computational grid

The computational grid is essential for the CFD case. Structured grids are generated for the periodic hill cases because the geometries are simple and the structured grids can offer better grid quality and less memory cost. The basic grids are generated by controlling the node distribution on each edge to guarantee $y^+ < 1$ and wall orthogonality, as shown in Fig. 6 (a). It can be observed that there exist two instances of spatial unsmoothness marked by red dashed lines because of the joints between the sloped parts and the flat part. To verify the effect of the computation grid on the mean flow features, smoothed grids are generated based on the basic grids. The smoothing process solves the elliptic PDE by iterating certain steps. The smoothed grid is shown in Fig. 6 (b). The unsmoothness disappears, and the transition is more moderate.

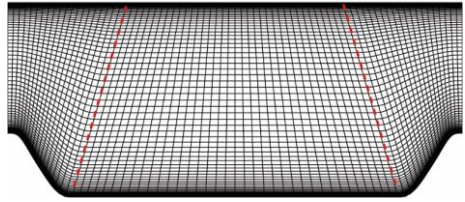

(a) The basic grid

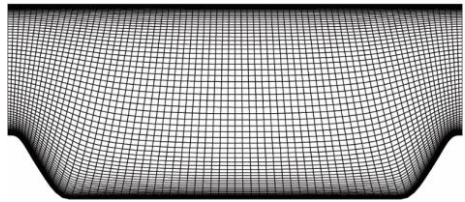

(b) The smoothed grid

Fig. 6 Computation grid of the periodic hill with $\alpha = 0.5$

The RANS simulations are performed on both sets of grids. The mean velocity profile comparison is shown in Fig. 7. The velocity profiles are very close, and no discontinuity exists in the basic grid result. The density and pressure also show the same phenomena, which indicates that the primitive variables are robust to minor adjustments. However, the input feature comparison shows the effect of grid smoothness. The input feature $q_6 = \mathrm{tr}\left(\hat{\mathbf{A}}_p^2 \hat{\mathbf{S}}^2\right)$ and $q_7 = \mathrm{tr}\left(\hat{\mathbf{A}}_k^2 \hat{\mathbf{S}}\right)$ distributions of the basic grid result and the smoothed grid result are shown in Fig. 8. These two features are both important in the machine learning model according to the following importance analysis. The distribution of the basic grid result (Fig. 8 (a) and (c)) shows a significant value jump marked by two black dashed lines, which correspond to the unsmoothness area of the computation grid (Fig. 6 (a)). The sudden decreases near the red lines certainly worsen the prediction capacity when applied on different cases with different grid distributions. In contrast, the distribution of the smoothed grid (Fig. 8 (b) and (d)) has no unphysical value jump. Compared with the primitive variables in the N-S equations, a portion of the input features is constructed based on the velocity gradient, pressure gradient, and TKE gradient. The derivative operation leads to a higher requirement for the solution smoothness and therefore needs better grid quality.

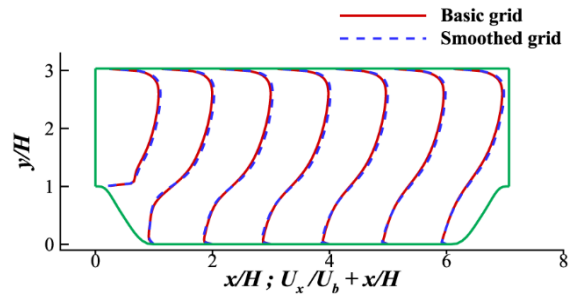

Fig. 7 Mean velocity comparison of the basic and smoothed grid

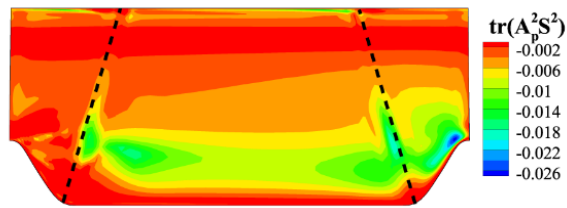

(a) $q_6$ on the basic grid

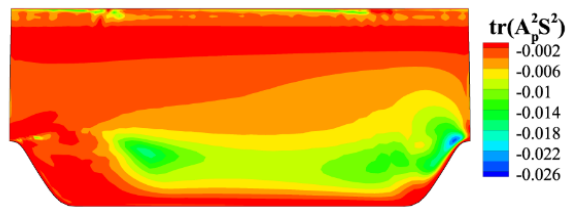

(b) $q_6$ on the smoothed grid

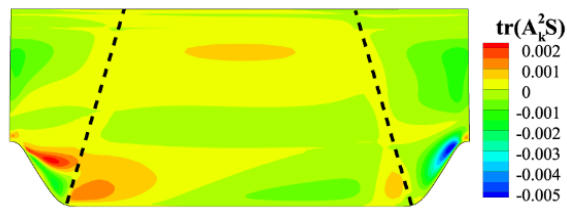

(c) $q_7$ on the basic grid

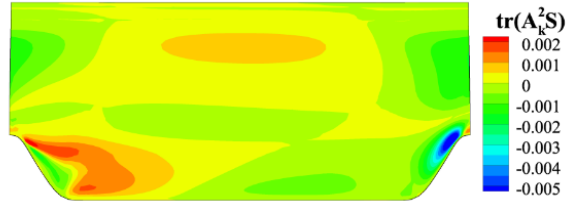

(d) $q_7$ on the smoothed grid

Fig. 8 Input feature comparison of the basic and smoothed grid for $\alpha = 0.5$

In summary, the machine learning framework put forward stricter requirements on grid smoothness. Different grids that have almost the same performance in the primitive variables may have a significant difference in the input features. The grids in the following research are all smoothed based on the same method.

## 2. The effect of the characteristic identification features

As illustrated in Sec. II B, the features based on characteristic identification analysis are added to the original tensor analysis-based input feature set to augment the relevance with the output targets. The feature importance analysis can help examine the effect of the added features. Based on a set of trained decision trees, the random forest can determine the importance of each input feature and therefore is employed in this work. The scikit-learn package in Python is used for programming [45]. The hyperparameters are set according to the work of Wu et al. [32]. The number of maximum features is set as 6, i.e., $1 + \log_2 n$, where $n = 25$ is the number of input features. The number of trees is set as 300. The training set is the same as the set used for the artificial neural network. Five random forests are trained separately for five output targets.

The top ten important features of each output target are listed in Table 1. The corresponding importance factors are also presented in parentheses. The importance ranking of the different targets vary, and the newly added features play important roles in all five output targets. Most features are within the top ten, or even top five, of the most important features, especially the ratio of the turbulent kinetic energy to the mean kinetic energy $f_4$, the wall-distance-based Reynolds number $f_6$, and the ratio of the turbulent viscosity to the mean flow viscosity $f_8$. The distributions of these three features are shown in Fig. 9. The three features separately focus on different flow areas and represent different characteristics. $f_4$ has a large value in the flow separation, giving out a marker for the area where the discrepancy mainly occurs. $f_6$ leaves out the boundary layers, guaranteeing the correct resolution of the flow near the walls. $f_8$ has a small value in the shear layer, locating the insufficient modeling. These features are closely related to the distributions of the output targets and therefore affect the machine learning framework.

In summary, the newly added features based on the characteristic identification work well in training and prediction and complement the connection between the input and output.

Table 1 The top ten important features of each output target measured by the random forest

| Output target | 1st | 2nd | 3rd | 4th | 5th | 6th | 7th | 8th | 9th | 10th |
|---|---|---|---|---|---|---|---|---|---|---|
| $\Delta \log k$ | $f_6$ (0.16) | $f_8$ (0.15) | $q_4$ (0.07) | $q_1$ (0.06) | $f_4$ (0.05) | $f_5$ (0.05) | $q_3$ (0.05) | $f_1$ (0.04) | $q_2$ (0.04) | $q_6$ (0.04) |
| $\Delta \xi$ | $f_8$ (0.13) | $f_6$ (0.10) | $f_4$ (0.09) | $q_7$ (0.06) | $f_1$ (0.05) | $q_{15}$ (0.05) | $f_2$ (0.05) | $q_3$ (0.04) | $f_5$ (0.03) | $q_9$ (0.03) |
| $\Delta \eta$ | $f_6$ (0.21) | $f_8$ (0.18) | $f_7$ (0.08) | $q_4$ (0.07) | $q_1$ (0.07) | $q_2$ (0.04) | $f_5$ (0.04) | $q_8$ (0.03) | $q_6$ (0.03) | $f_4$ (0.03) |
| $\cos\left(\Delta \theta_z^1\right)$ | $f_8$ (0.17) | $q_7$ (0.11) | $f_6$ (0.10) | $f_5$ (0.06) | $f_4$ (0.06) | $q_8$ (0.05) | $q_{11}$ (0.04) | $q_1$ (0.04) | $q_4$ (0.04) | $f_1$ (0.04) |
| $\mathrm{sgn}\left(\Delta \theta_z^1\right)$ | $f_8$ (0.15) | $f_4$ (0.10) | $q_3$ (0.08) | $q_{11}$ (0.07) | $q_7$ (0.06) | $q_8$ (0.06) | $f_6$ (0.05) | $q_{14}$ (0.05) | $f_1$ (0.04) | $q_1$ (0.03) |

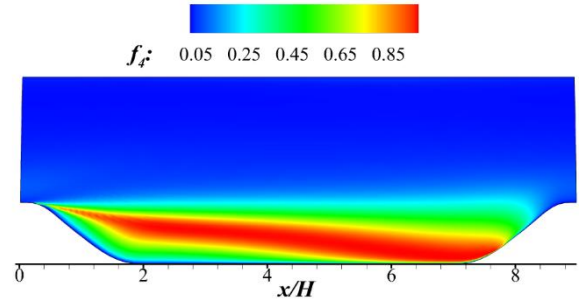

(a) The ratio of the turbulent kinetic energy to the mean kinetic energy $f_4$

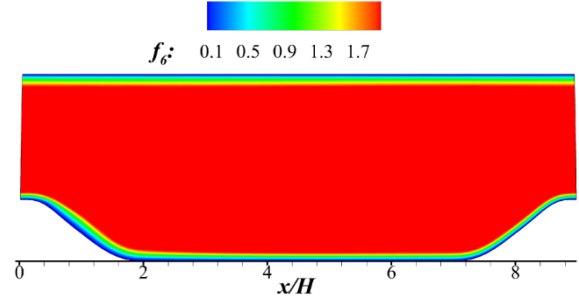

(b) The wall-distance-based turbulent Reynolds number $f_6$

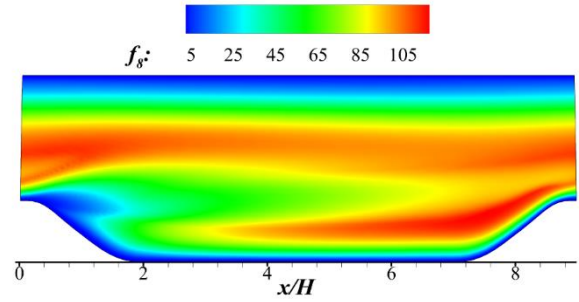

(c) The ratio of the turbulent viscosity to the mean flow viscosity $f_8$

Fig. 9 Representative features based on characteristic identification

### 3. The effect of improved representation of spatial orientation

As illustrated in Sec. II C, the original rotation angle distribution may have a numerical discontinuity problem. The $\Delta\theta_z^1$ distribution in the periodic hill case is shown in Fig. 10 (a). The areas where the value jumps from $\pi/2$ to $-\pi/2$ are marked by black boxes. When applying the spatial orientation decomposition, the $\cos\left(\Delta\theta_z^1\right)$ and $\mathrm{sgn}\left(\Delta\theta_z^1\right)$ features are separately shown in Fig. 10 (b) and (c), respectively. The original value jumps are transferred to the minimum value of $\cos\left(\Delta\theta_z^1\right)$ and the sign switching of $\mathrm{sgn}\left(\Delta\theta_z^1\right)$. The $\cos\left(\Delta\theta_z^1\right)$ distribution is moderate and suitable for the regression. The $\mathrm{sgn}\left(\Delta\theta_z^1\right)$ distribution shows the clustering characteristics. The positive and negative boundaries have a certain pattern, which can be learned by the neural network.

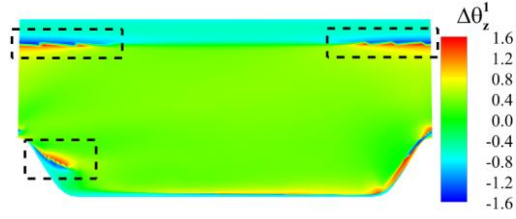

(a) The distribution of $\Delta\theta_z^{\mathrm{I}}$

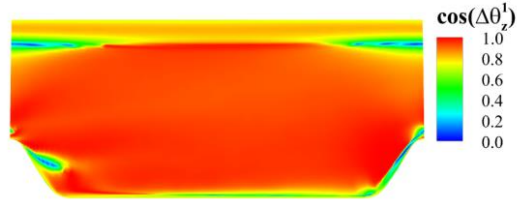

(b) The distribution of $\cos\left(\Delta\theta_z^{\mathrm{I}}\right)$

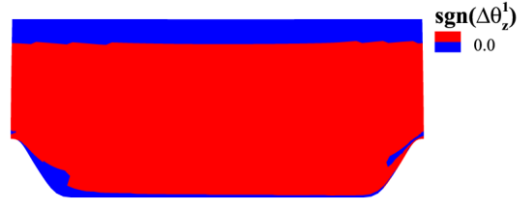

(c) The distribution of $\mathrm{sgn}\left(\Delta\theta_z^{\mathrm{I}}\right)$

Fig. 10 The original and decomposed spatial orientation distribution for $\alpha = 0.5$

## C. The generalization performance in varying geometries

Two sets of neural networks are trained. The first set is trained on the basic grids and the original spatial orientation representation, marked as "original". The second set employs the smoothed grids and the decomposed orientation features, marked as "modified". The other processes are kept the same. Each set of networks includes five networks for each output feature. As stated in Sec. II D, k-fold cross-validation is employed for parameter tuning. Because the train set consists of the cases $\alpha = 0.8$ and 1.0, the train set is separated into 2 folds. The adjustable hyperparameters of one neural network are shown in Fig. 11, including the number of hidden layers $L$, the node number of each hidden layer $N_i$ ($i$=1, 2, …, $L$), the dropout rate of each layer, the batch size, and the training epochs.

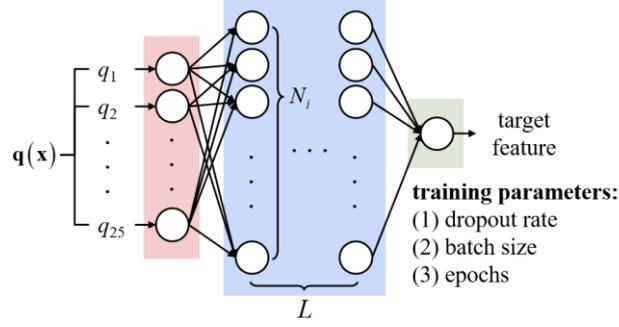

Fig. 11 The hyperparameters of the neural network

The rectified linear unit (ReLU) function is used for activation. The Adadelta algorithm is used for optimization. For the sign feature, the sigmoid function is used for the output layer activation, and binary cross-entropy is used as the loss function for classification. For the other features, the mean absolute error function is used as the loss function for regression. Because the cosine value is constrained to the closed set [0, 1], the sigmoid function is used as the activation in the output layer to guarantee the correct value range, while the other values employ linear functions. The optimized hyperparameters of the "modified" set of neural networks are shown in Table 2. Each feature has its own best hyperparameters. The neural networks for the TKE and the sign feature require fewer nodes and training epochs compared with other features, corresponding to their feature complexity.

Table 2 Optimized hyperparameters of the "modified" set of neural networks

| Output target | $L$ | $N_i$ | dropout rate | batch size | epochs |
|---|---|---|---|---|---|
| $\Delta \log k$ | 2 | [128, 64] | 0 | 512 | 4500 |
| $\Delta \xi$ | 3 | [128, 64, 32] | 0 | 512 | 6000 |
| $\Delta \eta$ | 3 | [128, 64, 32] | 0 | 512 | 6000 |
| $\cos\left(\Delta \theta_z^1\right)$ | 3 | [128, 64, 32] | 0 | 512 | 6000 |
| $\mathrm{sgn}\left(\Delta \theta_z^1\right)$ | 2 | [128, 64] | 0 | 512 | 3000 |

After the training process is completed, the two sets of neural networks are applied to predict the three test cases. Because of the strong fitting capacity, the results of the train set all have good consistency with the targets and therefore are not presented. The focus is on the test set. To verify the improvement of the modified method on the test cases, the predicted TKE profiles are compared in Fig. 12. The RANS results, DNS results, and prediction results of the original/modified method are presented. The modified method provides a more accurate prediction regardless of the interpolation ($\alpha = 1.0$) or the extrapolation cases ($\alpha = 0.5, 1.5$). Compared with the original method, the smoothness and the magnitude of key areas, such as the beginning of the shear layer, are both improved, which verifies the effect of approaches employed in this work.

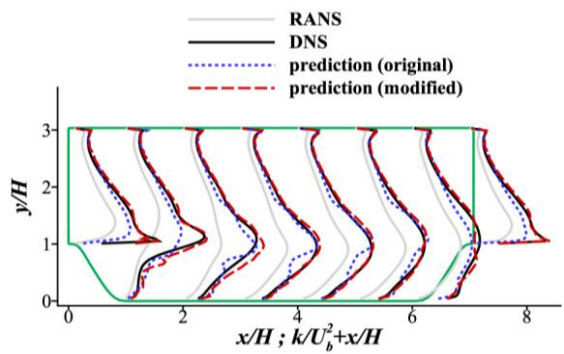

(a) The test case: $\alpha = 0.5$

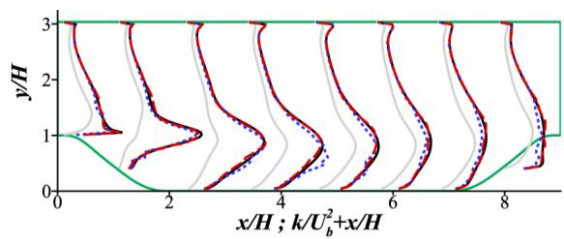

(b) The test case: $\alpha = 1.0$

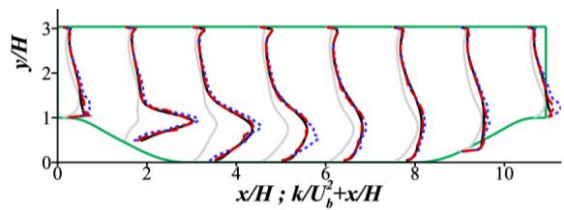

(c) The test case: $\alpha = 1.5$

Fig. 12 TKE profile comparison of test cases

Substituting the predicted Reynolds stress fields into the CFD solver and iterating until reconvergence, the modified mean flow fields are acquired. The residual convergence situations are shown in Fig. 13. The baseline RANS computation is performed and converged to provide an initial field. After substituting the Reynolds stress, the residual rises and then decreases again. Substituting the true Reynolds stress gives the least residual. Substituting the prediction results causes similar behavior but creates a higher residual level.

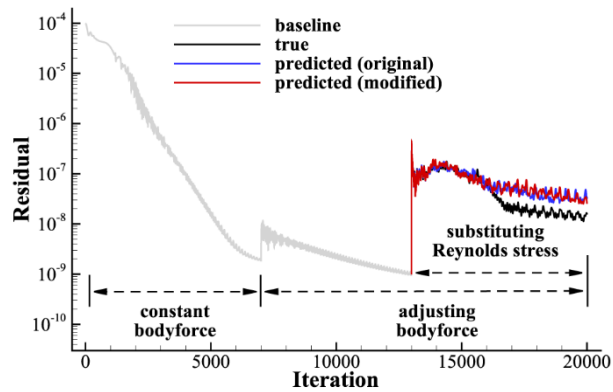

Fig. 13 Residual convergence history of the test case $\alpha = 1.0$

Till now, all the steps needed in the machine learning assisted framework are finished. To verify the feasibility of applying to practical turbulence problem, the time costs of the entire process are collected and list in Table 3. The first four steps only take place once and do not need to perform in each CFD solving because the stress is frozen during the iteration. It can be seen that the parameter tuning using the cross-validation takes the longest time. This is because lots of alternative hyperparameters are considered. After acquiring the optimal model parameters, the training process costs much less time and the time cost of other steps can basically be ignored. Therefore, the time cost of the current framework is suitable for the practical turbulence problem, especially for those optimization problem needing assessing lots of individuals quickly.

Table 3 Time costs of the entire machine learning assisted framework

| Steps | CPU time |
|---|---|
| Grid smoothing | **<1s** for one set of grid |
| Cross-validation | **~5200s** for each output target, **total 26000s** |
| Model training | **~160s** for each output target, **total 825s** |
| Model prediction | **<1s** for each output target |
| CFD solving | **basically the same** with SST converge cost |

The velocity contour comparison is shown in Fig. 14, taking the test case $\alpha = 1.0$ as an example. The velocity fields are all nondimensionalized by the bulk velocity at crest $U_b$. The DNS result has a smaller separation bubble compared with the RANS result, which results from the insufficient mixture modeling of the RANS simulation. This can also be observed from the magnitude of Reynolds stress in Fig. 12. The prediction results supplement the discrepancy of Reynolds stress and therefore reduce the separation bubble length. However, the result of the original method shows obvious unsmoothness in both the mainstream and separation areas. The phenomenon is more severe in the other two extrapolation cases. In contrast, the result of the modified method has better smoothness and is more consistent with the DNS result.

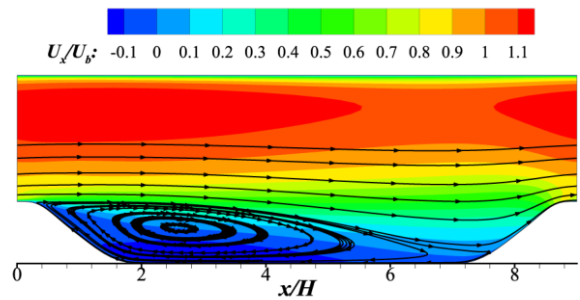

(a) RANS result

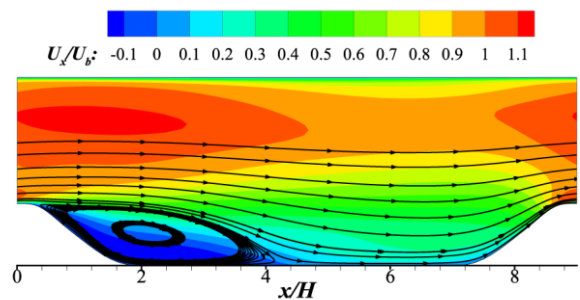

(b) DNS result

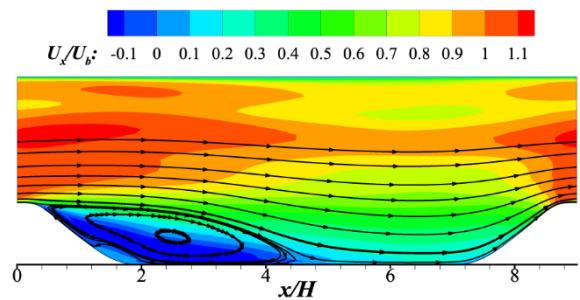

(c) Prediction result (original method)

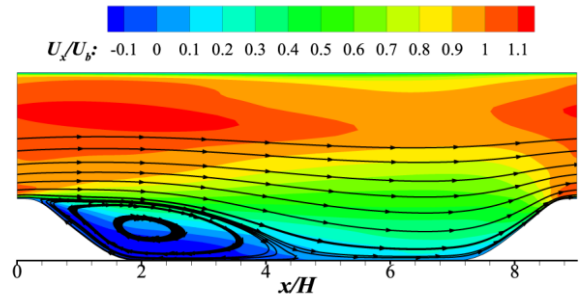

(d) Prediction result (modified method)

Fig. 14 Velocity contour comparison of the test case: $\alpha = 1.0$

To further illustrate the mean flow prediction performance quantitatively, the velocity profiles are extracted and compared in Fig. 15. The results of the modified method have better performance compared with the original method. The reattachment locations of the three cases are all improved significantly, and the entire velocity profiles are much closer to those of the DNS results than those of the RANS results. The difference is mainly located in the mainstream area. The unphysical inflections indicate that the mainstream flow is more sensitive to the Reynolds stress difference; this phenomenon also appears in the theoretical optimum result above (Fig. 5 (c)) and the research of Wu et al. [48].

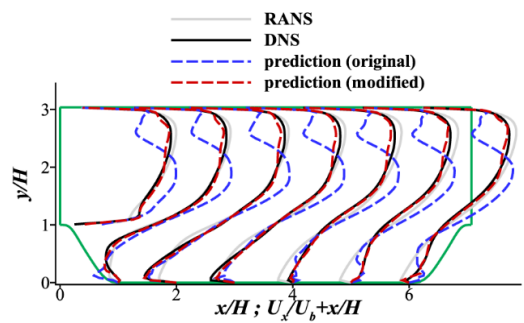

(a) The test case: $\alpha = 0.5$

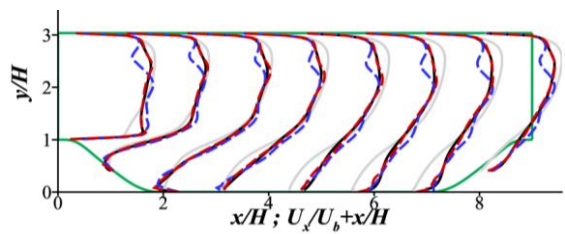

(b) The test case: $\alpha = 1.0$

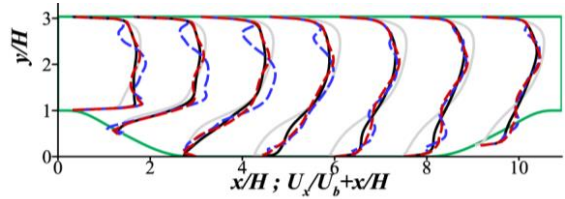

(c) The test case: $\alpha = 1.5$

Fig. 15 Velocity profile comparison of the test cases

The physical quantities located at the separation side are important in engineering applications. Therefore, the bottom wall of the periodic hill is further investigated. The surface friction coefficient distributions are shown in Fig. 16. Regardless of the test case, the friction distribution of the prediction result is significantly improved and closer to that of the DNS result. The shape of the former half curve corresponds to the velocity profile and the boundary layer development in the recirculation area. There are two key positions in the first half curve. The position where the friction becomes positive indicates the reattachment point. The turning point where the friction reaches a minimum before the reattachment indicates the strongest reverse velocity and illustrates the core of the separation area. These two positions are marked in Fig. 16, and the positions of the RANS results are not marked because of the large difference from the other two results. The key positions predicted by the machine learning framework are improved and close to the DNS results. The error of the test case $\alpha = 1.5$ is relatively large, but the curves show close patterns. The shape of the latter half of the curve demonstrates the flow situation after reattachment. There also exist two key locations with the maximum/minimum value. The first is located at the connection between the bottom and the hill where the flow area starts to contract. The sudden inflection leads to a decrease in friction. This phenomenon is hardly observed in the RANS results because the separation areas are too large and the inflection point is covered by the separation. In contrast, the prediction results are quite consistent with the DNS results in all three cases. The second corresponds to the peak value near the hillcrest, which is located at the hillside close to the crest. The DNS results have larger peak values and generally higher friction coefficient levels. This is because the Reynolds stress near the hillside of the DNS results is larger than that of the RANS results, leading to a stronger shear flow. The prediction results fit the true distribution quite well.

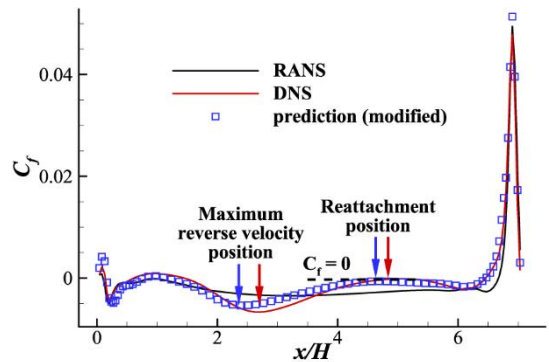

(a) The test case: $\alpha$ = 0.5

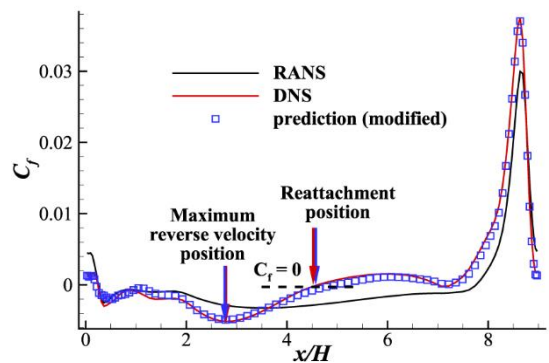

(b) The test case: $\alpha$ = 1.0

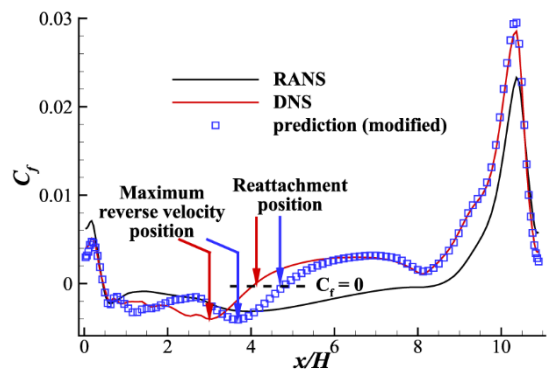

(c) The test case: $\alpha$ = 1.5

Fig. 16 Surface friction comparison at the bottom wall of the test cases

The pressure coefficient distributions are shown in Fig. 17. Compared with the friction distributions, the prediction results have better consistency with the DNS results. This is because the pressure distribution is closely related to the flow area of the mainstream and therefore is determined by the shape and size of the separation area. The separation areas of the prediction results are close to those of the DNS results, as shown in the velocity profile comparison in Fig. 15, and therefore lead to a better pressure prediction.

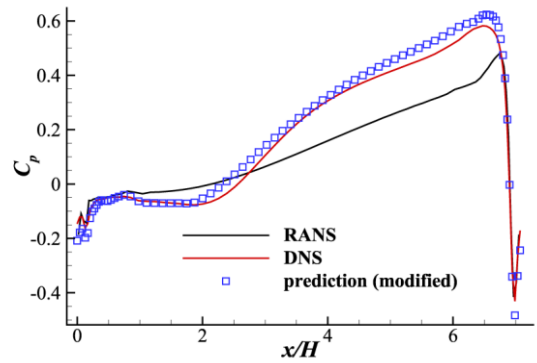

(a) The test case: $\alpha = 0.5$

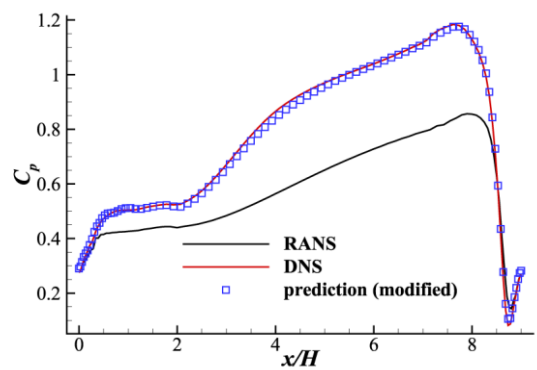

(b) The test case: $\alpha = 1.0$

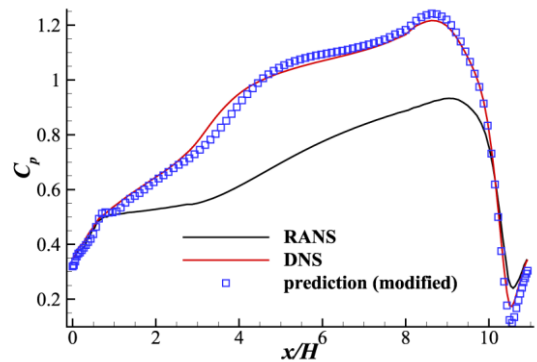

(c) The test case: $\alpha = 1.5$

Fig. 17 Pressure distribution comparison at the bottom wall of the test cases

In summary, the results presented above show the prediction abilities of the current framework. The modified input features, smoothed grids, and the employed spatial orientation decomposition further improve the generalization, extending the extrapolation to a wider varying geometry range.

## IV. Conclusion

Among the three research directions of the data-driven approaches, only machine learning-assisted turbulence modeling can theoretically predict the entire true value of Reynolds stress. The method with the most potential is accompanied by the most difficult modeling and training process. To further improve the performance of machine learning-assisted turbulence modeling, a modified machine learning framework, which focuses on solving the smoothness problem and enhancing generalization, is proposed in this work. The input feature selection criteria based on the physical and tensor analysis are proposed and employed to supplement several effective features. The intrinsic discontinuity problem of the spatial orientation difference is analyzed, and the feature decomposition method is introduced to reduce the prediction difficulty. The smoothness of the computation grid is investigated and shown to have a significant effect on the input features consisting of high-order derivatives of the primitive variables. Applying the modifications above, the neural networks are trained based on the periodic hill cases. The prediction results show better smoothness and accuracy. The generalization is proven to be valid on large geometric differences. After substituting the predicted Reynolds stress into the mean flow solution, the sizes of the separation bubbles, the reattachment positions, the pressure and the friction distributions on the separation sidewalls are corrected significantly compared with the original RANS results.

## Acknowledgments

This work was supported by the National Natural Science Foundation of China (91852108 and 11872230). The assistance of Prof. Heng Xiao and Dr. Jin-Long Wu is greatly appreciated.

# Data availability statements

The data that supports the findings of this study (the dataset of flows over periodic hills of parameterized geometries) are available within the article [46].